\documentclass[11pt]{article}

\makeatletter
\def\eqnarray{\stepcounter{equation}\let\@currentlabel=\theequation
\global\@eqnswtrue
\global\@eqcnt\z@\tabskip\@centering\let\\=\@eqncr
$$\halign to \displaywidth\bgroup\@eqnsel\hskip\@centering
  $\displaystyle\tabskip\z@{##}$&\global\@eqcnt\@ne
  \hfil${\;##\;}$\hfil
  &\global\@eqcnt\tw@ $\displaystyle\tabskip\z@{##}$\hfil
   \tabskip\@centering&\llap{##}\tabskip\z@\cr}
\makeatother

\newcommand{\df}{{\rm d}}
\newcommand{\ii}{{\rm i}}
\font\SYM=msbm10 
\def\Real{\hbox{\SYM R}}

\begin{document}
%
\title{On singularities, horizons, invariants, and the results of Antoci,
  Liebscher and Mihich (GRG 38, 15 (2006) and earlier)}
\author{Malcolm A.H. MacCallum\thanks{School of Mathematical Sciences,
     Queen Mary, University of London,
     Mile End Road,
     London E1 4NS,
     U.K.
Email: m.a.h.maccallum@qmul.ac.uk}}
\date{}
\maketitle
\begin{abstract}
  Antoci {\it et al.\/}\@ have argued that the horizons of the
  boost-rotation, Kerr and Schwarzschild solutions are singular,
  having shown that a certain invariantly-defined acceleration scalar
  blows up at the horizons. Their examples do not satisfy the usual
  definition of a singularity. It is argued that using the same term
  is seriously misleading and it is shown that such divergent
  functions are natural concomitants of regular horizons. In
  particular it is noted that the divergence is given by the special
  relativistic approximation to the overall metric.  Earlier work on
  characterization of horizons by invariants is revisited, a new
  invariant criterion for them is proposed, and the relation of the
  acceleration invariant to the Cartan invariants, which are finite at
  the horizons and completely determine the spacetimes, is examined
  for the C-metric, Kerr and Schwarzschild cases. An appendix
  considers coordinate identifications at axes and horizons.
\end{abstract} 
 
\section{Introduction}

In a recent paper Antoci {\it et al.\/}\@ \cite{ALM4} argue that the
bifurcate horizon in boost-rotation metrics is singular in an
invariant sense.  Their argument follows very closely one which they
used for the Kerr, Schwarzschild and other solutions in earlier papers
\cite{ALM1,ALM2,ALM3}, and which is also mentioned in an unpublished
preprint \cite{ALMU} and the editorial comments by two of them on
Schwarzschild's original paper \cite{AL}.  In \cite{ALM4} they
summarize it as proving that ``a local, invariant, intrinsic
singularity occurs when approaching the horizon''\footnote{It would
seem from the cited papers that this is only one of the reasons that
Antoci {\it et al.\/}\@ think that the Schwarzschild solution cannot,
or perhaps should not, be extended across the horizon, and the horizon
should be regarded as singular.  I will not go into all these other
arguments: one of them is touched on in the Appendix here.}.

The claim is based on the fact that in each case the acceleration of a
timelike unit vector field tangent to the orbits of a
uniquely-determined timelike Killing vector is divergent as the
horizon is approached. (Actually, the limits taken may approach only
the two-sphere where past and future horizons meet, as discussed in
the Appendix, but I ignore this criticism in the main text.)  This
divergence is well-known for the Schwarzschild case, was derived for
the Kerr-Newman metric \cite{ALM1} and the Zipoy-Voorhees or Bach and
Weyl `gamma' metric \cite{ALM2}, and calculated approximately for the
C-metric as a representative of the boost-rotation cases
\cite{ALM4}. It was also re-derived for the Schwarzschild case by
taking a two-body solution due to Bach and Weyl and considering the
limit as the mass of the second body tends to zero \cite{ALM3}.

Presence of a diverging invariant is not sufficient to prove the
presence of a singularity in the sense now standard in the field.  In
the following section I point out that the acceleration invariant of
Antoci {\it et al.\/}\@ does not produce a singularity in this sense,
and show, from the special relativistic approximation, that a
divergent acceleration is to be expected whenever one has a regular
horizon. Moreover it is noted that the approximate calculation of
Antoci {\it et al.\/}\@ in the C-metric is an exact calculation in the
manifestly non-singular Minkowski space.

Antoci {\it et al.\/}\@ \cite{ALMU} argue that the special
relativistic analogy is inappropriate since it is of crucial
importance that the vector field considered is invariantly defined and that
the acceleration is therefore referred to some standard of absolute
rest. I present below a contrary view.

The final section of this paper studies the relation of the
acceleration invariant to the Cartan invariants which characterize the
solutions but remain finite on the horizons. It has been known for
some time that in the Kruskal-Szekeres spacetime, which is the maximal
analytic extension of the Schwarzschild metric, the obvious scalar
polynomial formed from the first derivatives of the curvature,
$R^{abcd;e}R_{abcd;e}$, vanishes exactly at the horizon \cite{KLA}
(which illustrates the existence of invariants whose blowup does not
imply a singularity: one has only to take $1/R^{abcd;e}R_{abcd;e}$).
I examine the link between this observation and the divergent
acceleration of Antoci {\it et al.} and show that the latter is
related to division by an invariant whose vanishing characterizes the
horizon.

\section{Conceptual aspects}

To discuss singularities we need first to define them. Singularities
in general relativity are defined by reference to incomplete,
inextendible, causal geodesics.  The reasons for using this definition
were elegantly and entertainingly explained by Geroch through a
Galilean dialogue \cite{Geroch}. There has been much work on the
occurrence and possible natures of singularities thus defined. For an
account of this see \cite{TCE} and, for a more recent review on
causal structure and boundaries, \cite{JMMS}.

It is well-known that the horizons of the Schwarzschild, Kerr-Newman
and boost-rotation solutions do not fulfil this definition. In these
cases, the regions discussed by Antoci {\it et al.\/}\@ are isometric
to regions of a larger manifold including the horizon in which every
point of the horizon is regular (i.e.\ has a neighbourhood which is in
first approximation a neighbourhood of Minkowski space).  In the
Schwarzschild case this larger spacetime can be taken in the
well-known Kruskal-Szekeres form. Hence geodesics can be extended
through and beyond the horizon, so the horizon is not singular. (Of
course, there still could be, and in the Schwarzschild and Kerr cases
there are, singularities inside the horizon.)

Divergence of an invariant only implies a singularity if its blow up
is sufficient to prove that geodesics cannot be continued. Antoci {\it
et al.\/}\@ have not shown that their invariants do this, for the
cases with horizons, and the results referred to above show that it is
not the case. In my view, using the word `singularity' in this context
is likely to be misleading and possibly tendentious (in that it might
lead an unwary reader to believe that the presence of a singularity in
the usual sense has been proved).

For completeness, one should note that the cases considered by Antoci
{\it et al.\/}\@ in which the Kretschmann scalar blows up do fulfil
the criterion for singularities at the boundaries. For instance, the
general `gamma'-metric or Zipoy-Voorhees solutions give examples of
`directional singularities' which include singular points, rather than
just regular horizons (for fuller discussion of their singularity
structure, see e.g.\ \cite{T1,T2,KH} and literature cited therein).
However, the presence of these singularities is not proved by the
blow-up of the acceleration invariant: the Zipoy-Voorhees family
includes the Schwarzschild metric which is not singular at the
horizon. (It is therefore hardly surprising that the Kretschmann
scalar's behaviour is discontinuous as a function of the parameter
characterizing members of this family, as noted in \cite{ALM2}.)   I
do not discuss further the cases with an divergent Kretschmann scalar.

So far I have not addressed the question of whether the behaviour
found by Antoci {\it et al.\/}\@ should be regarded as singular in
some sense, regardless of its relation to the usual definition. They
remark \cite{ALM3}, apropos of singular acceleration in the
Schwarzschild case, that ``this singular behaviour of the
gravitational pull ... has not generally aroused very much
concern''. Taking a Newtonian analogy, as Antoci {\it et al.\/}\@ do
\cite{ALM1}, citing Abrams, divergent accelerations would suggest a
point source singularity.  The more appropriate analogy with special
relativity I now describe (and which is discussed in \cite{ALMU},
which its authors kindly drew to my attention on reciving an earlier
draft of this paper) shows that such divergent accelerations are
naturally associated with any regular horizon, despite Antoci {\it et
al.\/}'s implicit assumption to the contrary. There is thus no cause
for concern.

Consider particles in special relativity. We need only consider
worldlines with constant acceleration, since Killing vector
trajectories must have this property. A standard exercise in special
relativity shows that, in units such that $c=1$ and in
pseudo-Cartesian coordinates whose $x$-axis is aligned with the
acceleration, a particle initially at rest at $x=1/b$, $y=z=0$, and
subject to a constant acceleration $b$ has the trajectory described by
\begin{equation}
\label{SRparticle}
x = (1+b^2t^2)^{1/2}/b.
\end{equation}
This curve is asymptotic to the null hyperplane $x=t$. Its acceleration is
inversely proportional to the proper time, along the geodesic with the
same initial position and velocity, between the initial point and the
null plane.

The curve (\ref{SRparticle}) is a trajectory of the Killing vector
field corresponding to boosts in the $x$ direction about the origin.
If we took all the trajectories of this Killing vector field, it is
easy to see their accelerations become divergent as we take a sequence
of trajectories approaching $x=t$, for example by taking initial
points on $t=0$ with $x\rightarrow 0$ from above.

It is not surprising that the more abruptly one wants to accelerate a
particle to the speed of light, the larger the acceleration
needed\footnote{Strictly, of course, a massive body in relativity can
  travel at the speed of light only if it loses its rest mass.}. Even
in Newtonian theory this would be true: however, Newtonian theory
gives us no reason to consider such a set of curves. Special and
general relativity were introduced precisely to explain observations
which had no Newtonian explanation, so it is not surprising they lead
to new considerations.

The general relativistic behaviour and the manifestly non-singular
special relativistic behaviour are analogous in the neighbourhood of
any regular point, as a consequence of the principle of
equivalence. One can make this more explicit by expanding the metric,
at a point on the regular horizon, in Riemannian normal coordinates:
only the leading flat-space term contributes to the divergence of the
acceleration, the role of the lower-order curvature terms being only
to pick out the invariant vector field. Any null plane (or line) in
Minkowski space has a similar set of associated Killing vector
trajectories, so the analogy leads to the conclusion that any smooth
regular null surface in general relativity can be associated with a
set of curves of divergent acceleration. In particular, when we have
an invariantly-defined set of Killing vector trajectories with a
regular lightlike horizon, we would expect this horizon to be
accompanied by divergent acceleration of the invariantly-defined unit
vector field tangent to the Killing trajectories.

In the Schwarzschild, Kerr and boost-rotation spacetimes, the extended
manifolds clearly reveal the horizons to be lightlike. Therefore any
massive body which does not cross them has to be accelerated so that
its worldline is asymptotic to a null curve, and the special
relativistic analogy just developed applies.  Indeed, it is the
special relativistic contribution which Antoci {\it et al.\/}\@
calculate in \cite{ALM4}, since the approximation used in their
equation (10) to derive the divergence is just the specialization of
their (1) which gives flat space. Hence they are using the divergence
in non-singular Minkowski space in arguing that the horizons of the
C-metric are singular!

In \cite{ALMU} it is argued that the analogy is not appropriate
because the divergent acceleration in special relativity is not an
invariant. Correspondingly, the null plane asymptotic to
(\ref{SRparticle}) is not an invariantly-defined horizon. The Antoci
{\it et al.\/}\@ method, as given in their papers, is limited to cases
where there is an invariantly-defined unit timelike vector field. If
there is more than a one-parameter symmetry group, the argument based
on acceleration of a timelike Killing vector requires a unique choice
among the possible Killing vector fields e.g.\ by their asymptotic
properties, as exemplified in \cite{ALM1}. We could make the special
relativistic calculation (\ref{SRparticle}) invariant by cutting a
wedge out of Minkowski space and considering the unique timelike
Killing vector field whose trajectories do not cross the boundary of
the wedge, and with respect to which the wedge is static. (This is
somewhat artificial, but it does use a region corresponding to the
region $r>2m$ of the Schwarzschild solution.)  There is no reason to
think that the invariant character of the timelike vector fields
considered by Antoci {\it et al.\/}\@ affects the question of whether
there is a true singularity.

However, it is true that the divergent acceleration of the
uniquely-defined vector field characterizes the position of the
horizon. In the next section we show how this can be understood in
terms of vanishing of a curvature invariant at the horizon. So it is
worth considering the question of analogies a little further. An
analogy should not be rejected just because it does not capture all
the features of a situation: if it did, it would not be an analogy but
an identical model. But the special relativistic approximation is more
than just an analogy: the existence of such an local approximation
precisely characterizes the regularity of a point in general
relativity, and in this sense the diverging acceleration coupled with
finiteness of curvature invariants could be said to confirm the
regularity of a horizon rather than weakening the argument for it.

Antoci {\it et al.\/}\@ consider the divergence of acceleration
significant because it relates to an invariantly-defined notion of
staticity. They do not differentiate, in this argument, between cases
where the length of the static Killing vector is bounded away from
zero (for which their view seems to me more defensible) and those
where the length tends to zero, for which the boost-symmetric
interpretation seems more natural, modelled on the trajectories of
boost Killing vectors in Minkowki space (or, more generally, locally
boost-symmetric solutions \cite{SKMHH}). 

The physical significance of the curves discussed by Antoci {\it et
al.\/}\@ is doubtful in another way. They are test body tracks (though
found as limits of other curves in \cite{ALM3}), in that there are no
masses moving on them and perturbing the gravitational field, and the
test bodies are accelerated although no non-gravitational force to
supply the acceleration is modelled. The consideration of the limit of
the Bach-Weyl two-body solution in \cite{ALM3} does not in my view
outweigh this criticism (in this case the second 'body' is a
singularity and the non-gravitational force is supplied by a 'strut'
or 'rod' along the axis).

Thus a more physical discussion would imply perturbing the solutions,
and then one would lose the invariant definition of the timelike
congruence on which the argument depends. This limitation might imply
that the approach has nothing to say even about perturbed
Schwarzschild black holes, let alone other cases where what Antoci
{\it et al.\/}\@ call the `gravitational pull' lacks invariant
definition. However, one would expect to be able to generalize the
argument and, from the special relativistic analogy, to find an
divergent acceleration: the special relativistic argument above easily
extends to sets of curves of constant acceleration which are not
trajectories of the same Killing vector, and seems likely to extend
also to cases where the acceleration is not constant along the curves.
Alternatively, the approach described in the next section may provide
a better way to generalize the idea.

\section{Cartan invariants and the acceleration invariant}

It is known (see \cite{SKMHH}, chapter 9, for a review) that the
Cartan invariants completely characterize the spacetime locally (an
explicit reconstruction from these invariants was given for
Schwarzschild and some other solutions in \cite{KL}).  The completeness of
the Cartan invariants as a specification of the local geometry gives a
very precise form to the remarks of Synge on the lines of
`gravitational field = curvature of space-time', which Antoci {\it et
  al.\/}\@ call vague and appear to reject \cite{ALM2}. Properties of
the global structure or topology which do not have local consequences
are not captured by the Cartan invariants, but the former do not include
the acceleration invariants of Antoci {\it et al.}

In this section I will show how the divergent acceleration invariant
relates to the Cartan invariants for the Kerr, Schwarzschild and
C-metric cases, in which the Cartan invariants are all finite or zero
at the horizon (as, consequently, are all scalar polynomial invariants
in the curvature and its derivatives). My approach to this problem was
prompted by the vanishing of $R^{abcd;e}R_{abcd;e}$ at the
Schwarzschild horizon. It is not in general possible to isolate the
acceleration term at a general point, but it can be done for the Kerr
(and, a fortiori, Schwarzschild) cases. However, even in the C-metric
the effect can be isolated on the symmetry axis.

I will consider the Kerr solution in Boyer-Lindquist coordinates, with
the usual coordinates for Schwarzschild as a special case, and use the
coordinates of \cite{HT} for the C-metric (the latter are more
convenient in some respects than the coordinates of \cite{SKMHH},
Table 18.2, or the Weyl coordinates used in \cite{ALM4}). Thus the
metrics are
\begin{equation}
\label{Kerr}
\df s^2 = -\frac{S\Delta}{\Sigma^2} \df t^2  + \frac{\Sigma^2}{S}\left(
  \df \phi - \frac{2amr \df t}{\Sigma^2}\right)^2 \sin^2 \theta -
  \frac{S \df r^2}{\Delta} + S \df \theta^2,
\end{equation}
where $S=r^2+a^2 \cos^2 \theta$, $\Delta = r^2-2mr+a^2$ and $\Sigma =
S-2mr$, $m>0$ and $a$ being constants, and
\begin{equation}
\label{Cmetric}
\df s^2 = \frac{1}{A^2(x-y)^2}\left(f(y)\df t^2 - \frac{\df y^2}{f(y)}
+ f(x)\df \phi^2 + \frac{\df x^2}{f(x)}\right),
\end{equation}
where $f(\xi) = (1-\xi^2)(1+2Am\xi)$ and $m$ and $A$ are positive
constants satisfying $2mA > 1$.

The coordinate ranges will be restricted to a region outside the
horizon.  In (\ref{Kerr}) the region considered is where $r > r_0$,
$r_0$ being the larger of the roots of $\Delta =0$. This includes the
Schwarzschild case, $a=0$, where the region is $r>2m$. The timelike
Killing vector considered by Antoci {\it et al.\/}\@ is a multiple of
$(r_0^2+a^2)\partial_t+a\partial_\phi$.  In (\ref{Cmetric}), the
appropriate region, where $\partial_t$ is a timelike Killing vector,
is where $-1 \leq x \leq 1$ and $-1/2mA < y < -1$. The acceleration
horizon is where $y \rightarrow -1$, the black hole horizon is where $
y \rightarrow -1/2mA$, and $|x| \rightarrow 1$ gives portions of the
symmetry axis. I will not make calculations on the horizon itself, but
for brevity I use `at the horizon' to mean `as we approach the
horizon'.  As well as avoiding any need for coordinate
transformations, this is done because the horizon itself is a
hypersurface where invariants which are generically non-zero may
vanish, leading to indeterminacy, except by analytic continuation, of
some quantities which are elsewhere determinate. I also do not
consider the regions inside the horizons, although an analogous
analysis can be given, since the objective is to compare with the
acceleration invariant of Antoci {\it et al.\/}\@ and acceleration
would not be an appropriate term where the relevant Killing vector is
spacelike.

The calculations follow the general procedure described in
\cite{SKMHH}, chapter 9, utilizing the Newman-Penrose notation
extended to denote spinors totally symmetrized on all dashed and on
all undashed indices by, for instance,
\begin{equation}
\nabla \Psi_{20'} \equiv \nabla_{(A|X'} \Psi_{|BCDE)} o^A o^B o^C
\iota^D \iota^E \bar{o}^{X'},
\end{equation}
where $\{o^A,\,\iota^A\}$ are the usual basis spinors.  Note that the
value of each invariant at a given point of the manifold is
independent of the coordinates used.  The Cartan invariants for the
Schwarzschild solution have been given in earlier work \cite{KL,PRM},
and a general discussion of Petrov Type D vacua has been given in
\cite{CDV1,CDV2}. All type D vacua were explicitly classified, using
Cartan invariants, by {\AA}man \cite{JEA}.

At the first step, classifying the curvature, we find we have in each
case a vacuum metric of Petrov type D. Choosing the frame to bring the
curvature into canonical form, the only non-zero invariant is
$\Psi_2$.  In the Kerr case, $\Psi_2=m/(r+\ii a \cos \theta)^3$, and
in the C-metric, $\Psi_2=-mA^3(x-y)^3$.  So far the frame is fixed
only up to a boost and a spatial rotation, the invariances of a Petrov
type D Weyl tensor. Thus no timelike vector (within the plane of the
two principal null vectors) is picked out. We may incidentally note
that the Kretschmann scalar can only blow up if at least one of the
Cartan curvature invariants does so.

The next step is to take the first derivative of the curvature. In
both the Kerr and C-metric cases, the boost can be used to choose a
frame such that $\nabla \Psi_{20'} = 3\rho \Psi_2 = -\nabla
\Psi_{31'}=3\mu \Psi_2$. With this choice $\rho = \nabla
\Psi_{20'}/3\Psi_2$ is also an invariant and is real. The vanishing of the
invariantly-chosen $\nabla \Psi_{20'}$, or equivalently, in the
present cases, $\rho= \nabla \Psi_{20'}/3\Psi_2$, is the criterion for
a horizon in terms of invariants (at least for Petrov type D
spacetimes). It is natural in that on the horizon the outgoing null
vector must be surface-forming and non-expanding, i.e.\ have $\rho=0$,
since the horizon is a marginally trapped surface.

Since $\nabla \Psi_{20'}= -\nabla \Psi_{31'}$ are the only non-zero
first derivatives for Schwarzschild, it follows that in the
Schwarzschild case the vanishing of $R^{abcd;e}R_{abcd;e}$ is an
equivalent condition.  However in general type D metrics, $\nabla
\Psi_{21'} = 3\tau \Psi_2$ and $\nabla \Psi_{30'} = -3\pi \Psi_2$ will
also be non-zero at a general point, and remain finite at the horizon
(in Kerr they are zero at the poles).  This explains Skea's finding
that results similar to those of \cite{KLA} do not obtain for scalar
polynomials in other cases \cite{Skea}. Since it answers one of the
other arguments sometimes made about the Schwarzschild and Kerr
solutions, it is worth noting that the Cartan invariants' values show
that the only essential constants are $m$, and $m$ and $a$
respectively (in the usual notation).

When $\tau \neq 0$, it is always possible to use the so-far
undetermined 'spin' (the rotation in the plane perpendicular to the
principal null directions of the Weyl tensor) to adjust the argument
of $\tau$: this can be done so that $\tau=-\bar{\tau}=-\pi$ in Kerr
and $\tau=\bar{\tau}=-\pi$ in the C-metric. The resulting values are
then invariants (the moduli are invariant whatever the argument, of
course).

The criterion above for the horizon implies we can easily generate
invariants which are divergent at the horizon, the simplest being
$1/\nabla \Psi_{20'}$. Others are $\Psi_2/\nabla \Psi_{20'}$, or, when
$\nabla \Psi_{21'}\neq 0$, $\nabla \Psi_{21'}/\nabla \Psi_{20'}$.
My objective is to extract the particular invariant used by
Antoci {\it et al.}

The equality $\nabla \Psi_{20'}= -\nabla \Psi_{31'}$ implies that the
derivative of $\Psi_2$ is orthogonal to the timelike vector of the
orthonormal frame associated with the Newman-Penrose one in the usual
way (i.e.\ as in \cite{SKMHH}, (3.12)), so this vector is orthogonal
to the surfaces on which $\Psi_2$ is constant. It is invariantly
determined in the cases considered (and can be similarly fixed in all
type D vacua, except those with local boost isotropy), and when there
is a hypersurface-orthogonal Killing vector (the C-metric and
Schwarzschild cases) the two vectors will be parallel.  In the Kerr
solution, there is no hypersurface-orthogonal Killing vector, and the
correspondingly chosen timelike vector is parallel to the vector
$(r^2+a^2)\partial_t+a\partial_\phi$ which is asymptotic at the
horizon $r=r_0$ to the Killing vector field used by Antoci {\it et
  al.\/}\@ \cite{ALM1}.

The invariants from the curvature and its first derivative do not give
the timelike vector's derivatives, such as the acceleration which is
given by the combination of spin coefficients
$(\epsilon+\bar{\epsilon}+\gamma+\bar{\gamma})/\sqrt{2}$. Generic
arguments have placed an upper bound of three derivatives on the
classification process for type D vacua \cite{CDV1,CDV2}, but it is
known that at most two are actually needed \cite{JEA}. We thus expect
the acceleration's value to appear, in the Cartan scalars, in the
second derivatives.  Where this occurs is somewhat obscured in the GHP
formalism \cite{CDV1,CDV2} since the crucial spin coefficients are
hidden inside the differential operators. Using the Newman-Penrose
form, the reason that the divergence of $\epsilon+\bar{\epsilon}$
and $\gamma + \bar{\gamma}$ does not show up in divergence of any
Cartan invariants in the present cases is clear: they always arise in
products with the terms $\nabla \Psi_{20'}$ and $\nabla \Psi_{31'}$
which vanish at the horizon.

We have no reason to expect that the expressions for the Cartan
invariants can be algebraically solved for the spin coefficients in
general.  They would of course still be specified by the Cartan
invariants, but indirectly by specifying the basis vectors whose
commutators give the spin coefficients. However, for Petrov type D
spacetimes with $\nabla \Psi_{20'} = \nabla \Psi_{31'}$ and $\nabla
\Psi_{21'} = -\nabla \Psi_{30'}$, there is a combination of the second
derivative invariants which involves the acceleration. Using the
relations found for the first derivatives,
\begin{eqnarray}
Q &\equiv& 2\nabla^2\Psi_{31'} +3(\nabla^2\Psi_{42'} -\nabla^2\Psi_{40'}
 - \nabla^2\Psi_{22'} + \nabla^2\Psi_{20'})/4  \nonumber \\[-8pt]
\label{Qdef}\\[-8pt]
&=&
 ((\beta+\bar{\beta})-(\alpha+\bar{\alpha}+\pi+\bar{\pi}))\nabla\Psi_{21'}
-(\gamma +\bar{\gamma}+ \epsilon+\bar{\epsilon}+2\rho)
    \nabla \Psi_{20'}. \nonumber
\end{eqnarray}
Note that specific values have not been substituted here: to do so
would obscure the logic because multiples of the same values may arise
in more than one way (e.g.\ both from partial derivatives and from
spin coefficients multiplying the terms differentiated).

In the Kerr case, $\alpha$, $\beta$ and $\pi$ are imaginary. Thus the
acceleration invariant is given by the combination of Cartan
invariants $Q/\nabla \Psi_{20'}+2\nabla \Psi_{20'}/3\Psi_2$. For the
C-metric there seems to be no such neat way to remove the
$\nabla\Psi_{21'}$ terms from $Q$ or any similar expression at a
generic point, without using values which cannot be directly
algebraically deduced from the Cartan invariants (e.g.\ the value of
$\beta+\bar{\beta}-\alpha-\bar{\alpha}$). It remains true, of course,
that $Q/\nabla \Psi_{20'}$ blows up at the horizon.

To directly obtain information about the acceleration from the Cartan
invariants in the C-metric, as distinct from inferring it from an
integration giving the tetrad commutators, we thus have to be more
subtle. The behaviour, if not the specific value, of the acceleration
we seek must be independent of $x$. The invariant $\nabla\Psi_{21'}$
must vanish at the axis of symmetry. So we could use the same
combination of invariants as in the Kerr case but evaluate it at or as
we approach\footnote{The wording here is chosen to cover the case
  where the axis itself is a conical singularity.  The relation
  between the interpretations of $f(x)=0$ as an axis and $f(y)=0$ as a
  horizon is explained by the discussion in the Appendix.} the
symmetry axis defined by $\nabla\Psi_{21'} \rightarrow 0$.  Note in
particular that in both this and the Kerr case, as we approach certain
points on the horizon all first derivative Cartan invariants vanish
but the second derivative ones do not.

Thus we have found an invariant criterion for locating the horizon, at
least in type D spacetimes, which reflects the fact that the horizon
is a non-expanding null surface. We have then shown that the
divergence of the acceleration at the horizon can be derived from
non-singular invariants by a division by the invariant which characterizes
the horizon by vanishing there. This entirely agrees with, and for these
cases gives precise expression to, the arguments of the previous
section for expecting a blow up in an acceleration invariant as the
horizon is approached.  This agreement, and the natural
interpretability of the criteria, gives strong reasons for expecting
the same ideas to carry over to any algebraically general or
non-vacuum solution with analogous horizon structure. The C-metric
case shows that these ideas are unlikely to be implementable in the
simple form of algebraic expressions for the acceleration in terms of
Cartan invariants at a general point.

\section*{Acknowledgements}
I am grateful to Profs. W.B. Bonnor and J.M.M. Senovilla for advice
and references, and to the principal authors of SHEEP/CLASSI (Inge Frick
and Jan {\AA}man) which was used to check some of the calculations.


\section*{Appendix: coordinates, axes and horizons}

Axes of symmetry and the horizons considered in this paper have in
common that they occur where, in some commonly-used coordinate chart,
a metric component tends to zero\footnote{The analogy is particularly
strong in the C-metric, where both part of the axis and the
acceleration horizon are at the zero $f(-1)=0$ of the $f$ in
(\ref{Cmetric}).} (and another to infinity). For brevity, I call the
system where this zero appears the `original system' of coordinates.
The axes and horizons in these cases can all be seen to consist of
regular points of spacetime by using an appropriate coordinate system,
which I will call the `regular system'.  However, if the original
system is considered to include its boundary, the transformation
between the systems usually becomes in some way pathological there,
though smooth at other points.  The pathology may be that some
function in the transformation blows up, or that the map is many-one.

That these pathologies do not in themselves make the change
unacceptable is easily seen from the example of the Euclidean plane,
taking polar coordinates as the original chart with metric
\begin{equation}
\label{polars}
\df s^2= \df r^2 + r^2 \, \df \theta^2,
\end{equation}
where $0 <r<\infty$ and $\theta$ and $\theta + 2\pi$ are identified
(or where $0 \leq \theta \leq 2\pi$). I will not bother about the
inconsistency that coordinate maps are required to be 1-1,
while the standard way of using $\theta $ is not, as this is easily
dealt with by using two charts, e.g.\ $-\pi < \theta < \pi$ and $0 <
\theta < 2\pi$, for the angular variable. Here $g_{\theta\theta}
\rightarrow 0$ as $r \rightarrow 0$. The transformation to Cartesians
(the regular chart), $x=r \cos \theta$, $y=r \sin \theta$ is many to
one if continued to $r=0$, since all values of $\theta$ are mapped to
the single point $x=y=0$. Nobody argues that this indicates anything
strange about the Euclidean plane.

One way of dealing with these apparently problematic situations
is to insist on the basic definition of a differential manifold, so
that coordinate chart neighbourhoods have to be open and their
boundary points are not included. (Recall that coordinate chart
neighbourhoods in a manifold are defined to be open sets in 1-1
correspondence with open sets in $\Real^n$.) Then the transformations
are not many-one and may contain divergent functions without any
departure from the usual manifold definitions, so nothing pathological
happens. The axis, or horizon, is then only covered by the regular
chart (which now has a perfectly satisfactory transformation to the
original chart at all points in their overlap). In the example above,
this means we stick to $r>0$ in (\ref{polars}), and only treat $x=0=y$
in Cartesians or other coordinates smoothly related to them at the
origin.

An alternative approach is to consider the need for compatibility of
the coordinate topology and the metric topology. (This compatibility
is mentioned in \cite{ALMU} but its implications are not fully
pursued.)  If we consider a manifold endowed in the usual way with a
smooth Riemannian metric, the two topologies agree on open sets,
including chart neighbourhoods. The problems mentioned above arise
when one tries to go from open chart neighbourhoods to neighbourhoods
with boundaries (or even corners).  For a manifold with boundary, the
coordinate neighbourhoods are open sets in the half-space, $H^n$ say,
and the boundary has (locally) the topology of $\Real^{n-1}$, from the
coordinate point of view (call this the `coordinate topology'). Trying
to make this extension, we find that for axes or horizons it is not
acceptable, because the two topologies no longer agree.

In the coordinate topology, for the Euclidean space example above, the
problem is seen when we assume the coordinate range of $r$ is $0 \leq
r < \infty$. What goes wrong at $r=0$ in the semi-infinite space $H^2$
in (\ref{polars}) is that the usual metric topology is no longer
Hausdorff: it is impossible to take two non-overlapping open sets, one
containing $(0,\,\theta_1)$ and the other containing $(0,\,\theta_1)$,
for $\theta_1 \neq \theta_2$. To make the topologies consistent we
have to identify all the points $(0,\, \theta)$, and the coordinate
transformation becomes many-one.

The metric topology for positive-definite Riemannian manifolds is well
understood, and it resolves the questions about axes (indeed can be
extended to take care of the cases where the axis is a conical
singularity with an angle defect, see e.g.\ \cite{HT}). In particular,
it deals with the C-metric at $|x| \rightarrow 1$, and the Kerr and
Schwarzschild solutions at the points where $\sin \theta = 0$.

The choice of an appropriate topology for indefinite (pseudo-)Riemannian
manifolds is less widely discussed, and indeed more than one has been
proposed \cite{HKM,TF}. However, they give the same conclusions for
the horizons under discussion. To be specific we could take the
Alexandrov topology. To illustrate the outcome, consider the
Schwarzschild solution in the form of (\ref{Kerr}) with $a=0$. If we
wish to take $r \geq 2m$ rather than $r > 2m$ we have to consider the
metric topology at $r=2m$ for all $t$, $\theta$, and $\phi$.

The consistency of topologies requirement is satisfied by the Kruskal-Szekeres
picture, in which different values of $\theta$ and $\phi$ at $r=2m$
remain distinct. This is consistent with the behaviour of the
three-dimensional positive-definite metric topology of the surfaces
$t=$constant. A few authors have argued that one should regard $r=2m$,
at a given $t$, as a single point: for this one has to violate the
metric topology in the opposite sense, unacceptably in my view, by
identifying points which appear to be finitely separated from one
another. The remaining issue is to consider whether different values
of $t$ at $r=2m$ should be identified. The transformation between the
Schwarzschild and Kruskal-Szekeres coordinates is
\begin{equation}
 \label{krusszek}
 u = (r/2m - 1)^{1/2}e^{r/4m}\cosh t/4m , \quad
 v = (r/2m - 1)^{1/2}e^{r/4m}\sinh t/4m.
\end{equation}
from which we easily see that for $r=2m$ and all finite $t$ we get
only the single sphere $u=v=0$. This is exactly what we would expect
from the analogy with behaviour at axes. Incidentally, we may note
that $r$ is a well-defined scalar function on the whole
Kruskal-Szekeres manifold (though not a well-defined coordinate since
its constant surfaces in general have two disconnected pieces).  The
future part of the horizon bounding the region we started with,
$0<u=v<\infty$, is the topological product of a sphere and a line, and
is the boundary $t \rightarrow \infty$ of the Schwarzschild coordinate
patch, and the past part $0<u=-v<\infty$ is similarly $t \rightarrow
-\infty$.  Thus the points of different finite $t$ with $r=2m$ have to
be identified, while the limiting `points' in the $(r,\,t)$ plane at
$|t|\rightarrow \infty$, $r \rightarrow 2m$, are lines. (The
calculations of Antoci {\it et al.\/}\@ in the usual
$(t,\,r,\,\theta,\,\phi)$ coordinates, if the limit is at finite $t$,
thus refer only to the sphere $u=v=0$ in the Kruskal picture: however,
there will still be a divergent acceleration at other points of $r=2m$
as argued above.)

Hence, it is hardly surprising that if one regards $r \geq 2m$ as the
coordinate patch, with the topology inherited from $H^n$, the
coordinate transformation at $r=2m$ to Kruskal or other coordinates in
which the metric is regular on the horizon is a singular one and not
1-1. However, this is no more objectionable than, and indeed directly
analogous to, transforming from polars to Cartesians at an axis in the
Euclidean plane. Similar remarks apply to the other regular horizons
considered.

\end{document}